%% file: main_arxiv.tex
\documentclass[conference,review]{IEEEtran} 
\IEEEoverridecommandlockouts

\usepackage{cite}
\usepackage{amsmath,amssymb,amsfonts}
\usepackage{algorithmic}
\usepackage{graphicx}
\usepackage{textcomp}
\usepackage{xcolor}
\def\BibTeX{{\rm B\kern-.05em{\sc i\kern-.025em b}\kern-.08em
    T\kern-.1667em\lower.7ex\hbox{E}\kern-.125emX}}

\usepackage{url}
\usepackage{microtype}
\usepackage{balance}
\usepackage{xfrac}

\input{PeterMacros}
\usepackage{tcolorbox}

\newcommand{\Meta}{Meta\xspace}
\newcommand{\SEV}{outage\xspace}
\newcommand{\SEVs}{outages\xspace}

\begin{document}

\title{Code Improvement Practices at \Meta}

\author{
\IEEEauthorblockN{
Audris Mockus\IEEEauthorrefmark{1}\IEEEauthorrefmark{2}, 
Peter C Rigby\IEEEauthorrefmark{1}\IEEEauthorrefmark{3},
Rui Abreu\IEEEauthorrefmark{1}, 
Anatoly Akkerman\IEEEauthorrefmark{1}, 
Yogesh Bhootada\IEEEauthorrefmark{1}, 
Payal Bhuptani\IEEEauthorrefmark{1}, \\
Gurnit Ghardhora\IEEEauthorrefmark{1}, 
Lan Hoang Dao\IEEEauthorrefmark{1}, 
Chris Hawley\IEEEauthorrefmark{1}, 
Renzhi He\IEEEauthorrefmark{1}, 
Sagar Krishnamoorthy\IEEEauthorrefmark{1}, 
Sergei Krauze\IEEEauthorrefmark{1}, \\
Jianmin Li\IEEEauthorrefmark{1}, 
Anton Lunov\IEEEauthorrefmark{1}, 
Dragos Martac\IEEEauthorrefmark{1}, 
Francois Morin\IEEEauthorrefmark{1}, 
Neil Mitchell\IEEEauthorrefmark{1}, 
Venus Montes\IEEEauthorrefmark{1}, 
Maher Saba\IEEEauthorrefmark{1}, \\
Matt Steiner\IEEEauthorrefmark{1}, 
Andrea Valori\IEEEauthorrefmark{1}, 
Shanchao Wang\IEEEauthorrefmark{1}, and 
Nachiappan Nagappan\IEEEauthorrefmark{1}}
\IEEEauthorblockA{\IEEEauthorrefmark{1}Meta Platforms, Inc.}
\IEEEauthorblockA{\IEEEauthorrefmark{2}The University of Tennessee, Knoxville}
\IEEEauthorblockA{\IEEEauthorrefmark{3}Concordia University, Montreal}
}
\maketitle
\begin{abstract}
The focus on rapid software delivery inevitably results in the accumulation of technical debt, which, in turn, affects quality and slows future development. 
Yet, companies with a long history of rapid delivery exist. Our primary aim is to discover how such companies manage to keep their codebases maintainable. 
{\em Method:} we investigate such company's (\Meta) practices by collaborating with engineers on code quality (via action research) and by analyzing rich source code change history
using mixed-methods to reveal a range of practices used for continual improvement of the codebase. We were also able to replicate aspects of prior industry cases studies on code reengineering.
{\em Results:} Code improvements at \Meta range from completely organic grass-roots done at the initiative of individual engineers, to 
regularly blocked time and engagement via gamification of Better Engineering (BE) work, to major explicit initiatives aimed at reengineering the complex
parts of the codebase or deleting accumulations of dead code. Over 14\% of changes are explicitly devoted to code improvement and the developers are given ``badges'' to acknowledge the type of work and the amount of effort. Our investigation to prioritize which parts of the codebase to improve lead to the development of metrics to guide this decision making. Our analysis of the impact of reengineering activities revealed substantial improvements in quality and speed as well as a reduction in code complexity.
Overall, such continual improvement is an effective way to develop software with rapid releases, while maintaining high quality.
\end{abstract}

\section{Introduction}

Over time, the codebase increases in complexity due to evolution in the functionality, ongoing maintenance, and developer churn~\cite{eick2001does}. It accumulates technical debt via design decisions that often focus more on the need to resolve issues fast instead of ensuring long-term maintainability~\cite{tom2013exploration}. Similarly, new developers may not be familiar with original design decisions~\cite{HM03b} and, by making incompatible changes, complicate future maintenance. Focus on rapid delivery, such as continuous integration and deployment~\cite{fitzgerald2014continuous}, place even more emphasis on speed 

~\cite{taplin2017move}. The presence of these -- and similar factors -- accelerates the degradation of the codebase, resulting in a tangled web of hacks, workarounds, dead code, and unfinished tasks that ultimately make source code more difficult to maintain. This, in turn, should dramatically slow down the development, defeating the original purpose of rapid delivery. While the practice of rapid delivery is used at \Meta, this predicament is avoided. We aim to discover what slows down, prevents, or reverses such natural code decay by studying software development practices at \Meta. 

While a substantial literature exists on how to improve software processes, few studies attempt to discover or catalogue a broad range code improvement practices in a large software company practicing rapid delivery. To conduct our study, we employ several approaches. First, we directly participate with teams doing code improvement initiatives. Second, we search the internal documentation, version control, and issue tracking artifacts for improvement activities. We, therefore, ask \textbf{RQ1:} what types of code improvement practices and activities exist in a large software company? If we discover these practices, it is still not clear how 
to prioritize them beyond general approaches of reducing the codebase, code complexity, or code smells. Therefore, \textbf{RQ2:} what information engineers need to prioritize major code improvement efforts?  
While it is relatively easy to identify major top-down code improvement activities, it may be much harder to detect efforts that emanate from individual engineers (if such exist). Hence, \textbf{RQ3:} do decentralized (and undocumented) bottom-up/organic code improvement efforts exist in a large company? Once we have answered the previous questions, it is not clear \textbf{RQ4:} what is the impact of the code improvement in terms of quality, productivity, lead time, and code centrality? To answer this, we replicate several aspects of previous empirical studies investigating the impact of reengineering~\cite{geppert2005refactoring,moser2007case,kim2012field}.

We discovered bottom-up approaches by analyzing the text
of commit messages, we participated in the strategic code
improvement initiative by helping identify the most central parts
of the codebase, we analyzed BE efforts by discovering past BE
activities recorded in the issue tracking data, and we replicated
a previous industry study on the impact of reengineering based
on the outcomes for explicit reengineering tasks.

Our main contributions include the discovery of policies, practices, and initiatives that appear to counteract natural code decay. Specifically, 
we found that a substantial portion of such activities were completely organic based on the initiative of individual developers. We also observed more structured practices, internally referred to as Better Engineering (BE), where some tooling, including gamification, were used to support and engage engineers to help with coding, documentation, and code cleanup activities. Finally, highly structured efforts with dedicated expert engineers were focused on improving strategically important parts of the codebase.

Our active participation in the strategic code improvement initiative to help identify the most central parts of the codebase has revealed approximately 20 criteria (some not previously reported in the literature) that may play a role in making major investments in code improvement.
Finally, our evaluation of the impact the code improvement brings in terms of quality, productivity, lead time, and code centrality suggests that, as in prior industry studies, the return on investment can be substantial. In contrast to prior studies that report relatively low percentage of effort for code improvement (approximately 4\%), we find 
a much larger portion (over 10\%) of code changes to be associated with some form of code improvement. 

We discuss the background on code decay (and improvement) studies in Section~\ref{s:back}, elaborate our contributions in Section~\ref{s:contr},
describe the methods and 
describe BE practices and tooling in Section~\ref{s:be}. We then turn to prioritization of code improvement efforts in Section~\ref{s:target} evaluation of the impact in
Section~\ref{s:eval} and discussion in Section~\ref{s:disc}. We conclude with this study's limitations in Section~\ref{s:limit} and conclusions in ~\ref{s:conc}. 

\section{Background and Literature}\label{s:back}

\textbf{Code Decay and Technical Debt.}
The observation that software becomes harder to maintain over time is quite old. For example, almost quarter century ago the code decay~\cite{eick2001does} phenomena was defined as ``code is more difficult to change than it should be'' and introduced indicators such as excessively bloated code, a history of frequent changes and faults, widely dispersed changes, kludges, and numerous interfaces to measure code decay. That work also discussed causes for software decay that are as relevant today,  including inappropriate architecture, violations of the original design principles, imprecise or changing requirements, time pressure, inadequate programming tools, organizational environment, programmer variability, and inadequate change processes with bad project management.  

Technical debt~\cite{tom2013exploration} is a similar but broader and more recent concept defined in Wikipedia as ``technical debt (also known as design debt or code debt) is the implied cost of future reworking required when choosing an easy but limited solution instead of a better approach that could take more time.''  Tom \textit{et al.}~\cite{tom2013exploration} describe several types of technical debt. Code debt which appears to be similar to code decay, e,g., ``unnecessary code duplication and complexity, bad style that reduces the readability of code, and poorly organised logic that makes it easy for a software solution to break when updated at a future point in time.'' The second type is design and architectural debt, such as, ``piecemeal design with an absence of reengineering.'' The third type is environmental debt related to development-related processes, hardware, and other infrastructure and supporting applications. Knowledge distribution and documentation is the fourth type of debt related to developer churn without adequate knowledge transfer, see, e.g.,~\cite{rigby2016quantifying}. The fifth type is testing debt manifested as lack of coverage or automation.

This literature describes numerous indicators of code decay, but lacks a unifying supply chain perspective 
that helps to deliberately investigate if the provided indicator set is complete and what part or type of supply chain is 
involved. We use software supply chain concept and operationalization to combine dependency-based (call), logical (co-change), and 
knowledge (authoring and reviewing) supply changes into a single network. 

Furthermore,   
\textbf{how to reverse code decay?}
What practices can counteract negative effects of code decay and technical debt? One of the more commonly used techniques is code refactoring or, more precisely, reengineering (the term refactoring ``transforming code without modifying semantics'' is grossly misused by software developers and even in academic literature, e.g., many transformations in the classical book on refactoring~\cite{fowler2018refactoring} are not semantics preserving). We thus use a more general (and accurate) term ``reengineering.'' In cases where software is not feature-complete and needs active maintenance or, especially, if it needs to accommodate features that were not considered in the original architecture, it may make sense to invest significant development effort to reengineer these parts of the codebase to make it easier to accommodate hitherto not contemplated features and, possibly, to reduce the technical debt accumulated over the years. 

Despite the widespread acceptance of the phenomena such as code decay and technical debt, surprisingly few studies investigate \textbf{the impact of reengineering activities.} For example, a large survey at Microsoft~\cite{kim2012field} found that engineers understand refactoring in much broader terms than simply semantics preserving code transformation (hence our use of the term ``reengineering'') and that reengineering entails substantial costs and risks~\cite{bavota2012does}. The investigation found that the most reengineered binaries had reduced numbers of dependencies and reduced number of faults. Improvements in productivity and quality have been documented in earlier studies of reengineering, e.g.,~\cite{geppert2005refactoring}. Reengineering may result in a more transparent codebase where it is easier not to overlook some unanticipated effects of a code change. Specifically, the study in~\cite{geppert2005refactoring} looked at reengineering 30 KLOC C$++$, ASN.1 generated code from a 3rd party protocol stack within a 7 MLOC system. The relevant part was modified by 40 different developers over five years. The effort resulted in virtual elimination of defects reported by end users, halving in the number of lines used to implement exactly the same functionality, and reduced the effort to implement an MR (an equivalent of a pull request) by 11\%. A later study in~\cite{moser2007case} found a similar result.

Most other studies of reengineering look at code metrics before and after reengineering, such as reduction in code smells~\cite{al2017empirical}, but the improvements in the code smell metrics may not correlate with any reductions in effort once reductions in total code size are accounted for~\cite{sjoberg2012quantifying}. Our contribution consists of cataloguing and reporting a variety of code improvement practices in a large company. 

Once code decay is identified and suitable approaches to remediate it are chosen, the question on \textbf{how to prioritize the remediation 
efforts} remains. Specifically, engineer-driven efforts may focus on parts of the system they are familiar with and where they do not need to coordinate their changes with numerous other developers. A study of Java projects~\cite{silva2016we} found that developers most frequently cited changing requirements as the main motivation for refactoring and an even larger study~\cite{pantiuchina2020developers} confirmed that structural metrics including code smells do not play a significant role in refactoring decisions. 
The most reengineering benefits may, however, come from reengineering the most complex parts of the system where  numerous engineers, often from different organizations, are actively making changes.  Hence some of the reengineering work may need to be done in a top-down manner with a careful evaluation of the potential risks and rewards. Our contribution is to develop multiple criteria and associated operationalizations to help prioritize code improvement efforts. 

Finally, \textbf{what can we expect as a result of code improvement methods?} Very few company studies exist, e.g.,~\cite{geppert2005refactoring,moser2007case,kim2012field} and each may have been affected by the particular organizational context. Replications are essential to  establish the generality of the findings in different contexts~\cite{wieringa2015six, gonzalez2012reproducibility, mockus2010experiences, rodriguez2018reproducibility,shackleton2023dead}. We, therefore, conduct a replication of the study by Geppert \textit{et al.}~\cite{geppert2005refactoring}.

\section{Contributions}\label{s:contr}

First, we discover and catalogue a wide range of code improvement efforts within a large company, \Meta. Previous catalogs have focused on identifying problem areas (e.g.,~\cite{hackbarth2010assessing}), developer perceptions (e.g.,~\cite{smith2016beliefs}), or experiences with agile development (e.g.,~\cite{leffingwell2007scaling}) at a high level. Our contribution lies in the methods used to discover these efforts and the resulting catalog of micro-practices that span from organic, developer-driven initiatives to top-down strategic efforts, as well as policies and tools that support these practices and increase engagement levels.
Second, we directly participate in and steer some of the code improvement initiatives. 
Third, we investigate an important question of how best to allocate the reengineering effort. Previous work primarily focused on code smells or code complexity, e.g.,~\cite{malhotra2015prioritization}, but such narrow focus does not capture the variety of criteria and constraints extant in large software organizations.
Fourth, we replicate and extend many elements of the prior studies investigating the impact of reengineering~\cite{geppert2005refactoring,moser2007case,kim2012field}. 
Fifth, we use the software supply network for measures of combined call graph, co-change, and authorship centrality as well as direct measures of productivity, development interval, and outages to help engineers prioritize which parts of the code need more attention.

\section{Context: software development at \Meta}\label{s:proc}

We first describe the key aspects of the software development process and tools needed to explain the data acquisition and analysis presented later.
At a high-level, \Meta has a company-wide search engine that, as regular internet search engines, allows search for any keywords. Second, documentation is tracked in online documents, 
and company-wide wikis, as well as interactive training tutorials. Fourth, the same systems for tracking tasks, issues, and code changes are used company-wide. Fifth, data associated with the usage of these tools is tracked in a data warehouse that has SQL access. 

At \Meta, we develop software for both our servers and client devices, including specialized 
hardware devices. This approach allows us to have fine-grained control over versioning and
configurations, and enables us to quickly push new code updates to production. Before any code 
is deployed, it undergoes rigorous testing, including peer review, in-house user testing, 
automated tests, and canary tests. Once the code is deployed, engineers closely monitor logs to 
identify potential issues.

At \Meta, we place a strong emphasis on code reviews as part of our development process. In addition to using IDEs, such as VS Code, version control system (mercurial) and numerous other testing and deployment tools, we use
Phabricator\footnote{\url{http://phabricator.org}} as the cornerstone of our CI system, which facilitates modern code reviews. Developers
submit their code for review, creating a patch representing the initial version of the code.
Reviewers can suggest improvements, leading to additional revisions until the diff is either 
approved and incorporated into the codebase or rejected. This process promotes high coding 
standards, helps detect flaws, and spreads knowledge throughout the organization.

In addition to our focus on code reviews and testing, we also have a formal process for reporting 
and addressing bugs, outages, or incidents.

By having a clear process for reporting and
addressing these issues, we can ensure that we maintain the quality and reliability of our products.

\section{Code improvement practices at \Meta}\label{s:be}

To investigate company-wide code improvement initiatives we gather information from internal company  documents that include meeting notes, planning documents, tutorials, wikis, tasks, diffs (at \Meta the term diff is used to refer to what is typically referred as a  pull request, i.e., logical change to one or more files that can be reviewed, has a test plane, etc.) , and other documentation. We start from keyword ``improving code'' and retrieve 
and inspect 20 top results. We then inspect the resulting documents for indications if they are describing a code improvement tool or practice
and use them to formulate additional keywords, such as, ``better engineering,'' ``code quality,'' ``code health,'' ``code coverage,'' ``code complexity,'' and ``refactoring,'' The procedure is then repeated for each of these subsequent keywords. We reached saturation and did not find new keywords that identified interesting documents and code.

To discover bottom-up efforts we search for specific diff tags and also for keywords in the title. Code improvement is ``perfective'' maintenance: ``code changes made with intention to make future changes easier''~\cite{swanson1976dimensions}. We, therefore, started from keywords associated with it in~\cite{mockus2000identifying}, such as ``cleanup'', ``unneeded'', ``remove'', ``rework''. We also added synonyms, such as ``delete'', more modern keywords such as ``refactoring'' and ``dead code'' as well company-specific terms such as
``better engineering'' and ``BE.'' We then sampled at least 20 diffs that contain these keywords as tags or in their title and refined the search to exclude occasional enhancements, like ``add deleted code.'' Finally, we investigated a mirror-image of code improvement, the so-called self admitted technical debt~\cite{potdar2014exploratory} by using keywords from that paper.

\subsection{Results for RQ1}

We discovered an internal course, wikis, announcements for BE events, and tools to support and measure code improvement and also to help increase engineer engagement for it. The topics included topics related not only to coding style and code complexity, but also to code reviews, testing and code coverage, code documentation. In addition to best practices, we found a number of tools to support, measure, and report 
code improvement activities. Some of the tooling was explicitly targeting engineer engagement, such as profile badges for reviewing diffs, reporting bugs, having high code coverage and so on. In addition to the best-practices document, we also found larger initiatives, 
such as Better Engineering (BE). The following is a brief summary of BE at \Meta. 

\begin{figure}[ht]

    \begin{centering}
    \includegraphics[width=0.4\textwidth]{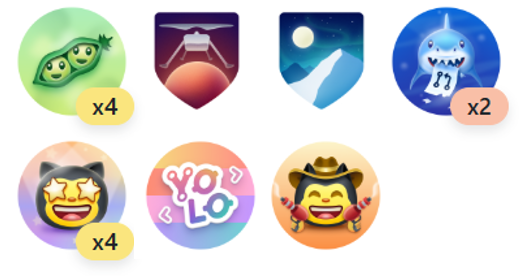}
   
    \end{centering}

    \caption{An engineer's profile may reach 70 or more badges. The images of badges shown internally are replaced with the analogues ``Achievement'' badges from GitHub. The set of internal badges is quite rich and includes badges such as ``Better Engineering (BE) gold'':  ``completed 20 tasks for High-Priority issue types'' and ``Code Cleaner --- Purple: deleted 100,000 or more lines of code''}
    \label{fig:badges}

\end{figure}

Better Engineering at \Meta is a company-wide initiative started in 2016 to improve engineering productivity in different codebases and tools.  It drives improvements in tooling, codebases and culture to improve productivity and engineering efficiency. The size and complexity of {\Meta}s codebases are increasing dramatically as it rapidly grows the worldwide engineering team, and continuous investment in this area is needed to keep the teams productive. Better Engineering is designed to improve engineering productivity across \Meta by keeping code modern with consistent abstractions and frameworks. It also creates a culture that recognizes and fixes engineering issues that slow teams down. It helps \Meta improve and optimize the developer experience and builds a sense of pride around the code and tooling. In general, \Meta's guideline is that teams should allocate 20\% to 30\% of engineering effort to BE projects. The work on BE is important and is recognized. BE is also supported by training, tooling and even gamification elements with profile badges (See Figure~\ref{fig:badges}) and scoreboards to encourage (and reward) participation~\cite{hamari2014does,seaborn2015gamification}. Gamification (drawing from game design) is the addition of gamified elements to a system such as badges or levels. A related concept of a public profile or  ``the quantified self~\cite{lupton2016quantified},'' which draws from wearables in the health realm and big data in the business realm, allow public display of various accomplishments. 

\begin{table}[ht]
\caption{Percent of diffs captured by each keyword. There is some overlap in selected diffs using different keywords 
as the percentages sum up to 16.5\% while the total percentage of diffs affected by any of the keywords is 14.2\%.}\label{t:perfective}

\begin{center}
\begin{tabular}{cr}
Keyword & \% of the diffs selected\\\hline
remove, delete, unneeded &7.26\% \\
better engineering	&	4.28\%\\
clean	& 3.01\%\\
refactor, rework	&	1.47\%\\
dead	&	0.52\%\\
\end{tabular}
\end{center}

\end{table}

The investigation of the ``perfective'' code change activity showed that a significant 14\% of diffs are devoted to perfective maintenance (see Table~\ref{t:perfective}. 
Tag or title bearing ``Better Engineering,'' is in 4.3\% of all diffs, ``refactoring'' in 1.5\%, and dead code removal in less than 1\% of the diffs, however. This suggests that most of the perfective maintenance is organic and is not a direct result of specific initiatives. 
Interestingly, despite the focus on rapid development, we found very little presence of self-admitted technical debt, with fewer than one in 10K diffs containing any such keywords.

In summary, we found code improvement activities to take many forms, ranging from major quality initiatives that are well documented,
to individual engineer driven actions that are less visible. 
Some of the code improvement initiatives target the strategically important parts of the codebase by producing a variety of indicators engineers could use to prioritize which parts of the codebase to target as described in Section~\ref{s:target}. 
In addition to courses, tutorials, extensive documentation, and regular scheduled code improvement activities, code improvement is  supported by tools that help track code metrics over time and also include various ways to engage developer 
via profile badges and even gamification of BE activities.

\section{Targeting Code Improvement Efforts}\label{s:target}

Reengineering may be costly in effort and may not always have a significant impact (or, at least, challenging to quantify). Furthermore, the effort that could be devoted to reengineering initiatives such as BE is inherently limited for actively maintained codebases. As such, it is critical to pick the reengineering tasks in a way to maximize the return on investment or to minimize risks and maximize rewards. Specifically, we are concerned with parts of the code that should be considered first when deciding on the scope and focus of the reengineering effort. Based on repeated interactions with the development teams, review of the related literature, and personal experience we proposed the following criteria for prioritization.

\subsection{Prioritization criteria}

While any code could be improved, the impact of that improvement may vary. Code decay and technical debt literature suggests  starting with the most decayed and fragile parts of the system, specifically, where changes are made by numerous authors, are difficult to complete, take a long time, and are likely to cause faults. 
Such areas may not be the first choice for organic individual developer-driven efforts due to the substantial investment of time and the need to coordinate changes to such parts of the code with multiple organizations.  
To find such areas we used a number of indicators proposed in the literature (see Section~\ref{s:back}) to measure decay and technical debt, with examples shown in Table~\ref{t:pri}. One novel indicator we also employ is the software supply chain (SSC) centrality~\cite{mockus2023modeling} as it appears to be strongly related to quality, lead-time, and productivity. 
Obviously, making any modifications to such parts of the code is a high-risk activity (we discuss risks below) and should also be considered in the prioritization.

\paragraph{Change activity}
As we observed in practice, the most decayed parts of the codebase may, however, be rarely or never changed. The potential benefits of being able to make changes faster, with less effort, and lesser chances of a fault are thus greatly diminished or eliminated (if no change will be needed in the future). Thus the total benefit of reengineering should take into account not just the effort savings for an individual change but also multiply it by the anticipated number of changes in the future.  
Actively changed files serve as a proxy of future effort and, even small improvements in such areas should yield significant overall results due to the frequency of changes. Reengineering of frequently changed code should yield much larger relative (per change) improvements to justify the reengineering effort and risk associated with reengineering. 

\paragraph{Knowledge loss}
It is important not to underestimate several hidden benefits of reengineering related to knowledge loss. Specifically, parts of the system originally designed and maintained by engineers who no longer work on the project carry significant risks~\cite{rigby2016quantifying}. Even a single change, if needed, may lead to serious issues and require significant effort. Reengineering such parts of the system makes current project members more aware of the specific design decisions needed to understand its operation and to simplify its maintenance.  
Each developer, by changing the codebase both learns from it and also imparts their own understanding. As such, parts of the code where a lot of changes were done by developers who left \Meta may not be well understood by anyone who is still with \Meta and require, if not reengineering, but at least a strong ownership. 

\paragraph{Authoring Speed}
A specific measure Diff Authoring Time (DAT~\cite{diff_authoring_time_podcast})\footnote{Gradle's DPE 2024 Summit \url{https://tinyurl.com/496225wf.}} was a preferred metric of development speed at the organization; it tries to capture exactly how much time does it take to author/write and land a diff. 
We, therefore, argued that in order to maximize the speed, the files that have the largest normalized DAT should be reengineered first.

\paragraph{Centrality}
Different parts of the codebase often vary in importance. The core or central functionality requires better quality controls and, generally, more attention. Hence, reengineering it may bring the most benefit. Previous work found that engineers perceive several distinct dimensions of code centrality~\cite{zhou2010developer} and several measures of code centrality were shown to be important predictors of engineer productivity~\cite{mockus2023modeling}. Code centrality can be calculated on a combined network of code dependencies, author-to-changed-file, and co-changes (all files changed in a single commit). These networks may also be considered separately resulting in multiple centrality measures, such as call-graph centrality and co-change centrality. For example, in a study of code decay~\cite{eick2001does} numerous interfaces was one of the indicators of decay, while a Microsoft study~\cite{kim2012field} considered the number of dependencies among binaries. Both of these represent ``degree centrality''  (the number of adjacent nodes), but we use Katz centrality, which also takes into account centrality of the adjacent nodes as well.
Specifically, we prioritize the parts of the codebase that are the most central with respect to SSC network, are actively changed (and are expected to be so in the future), have prior changes that lead to outages, and have experienced a significant knowledge loss~\cite{rigby2016quantifying}. Each dimension is described in more detail below.

\paragraph{Reliability}
At \Meta, all outages are investigated to identify their trigger. In case the trigger is a code change, the offending diff is noted in a special \SEV database. Outages are extremely infrequent hence only a tiny fraction of diffs associated with  changes are not typically marked as fixing a serious bug, but changes to address SEVs may be a way to assess some quality aspects of the code. Hence files that have high severity or large numbers of \SEVs fixed in them may need to be considered a priority.

\subsection{Results for RQ2}

For RQ2, we show the engineers the metrics and get their feedback on the importance.  Table~\ref{t:pri}, shows the definition operationalizes each of the criteria through multiple metrics. We calculated each of the metrics for the files that a team of engineers was responsible for and presented them in a spreadsheet and ordered them by centrality/importance. The engineers could resort to them based on other criteria and provide feedback on  the files. In the next section, we use these metrics to guide the reengineering efforts and measure the impact of the rework.

\begin{table*}[ht]

\caption{Various prioritization criteria presented with each candidate file. Engineers may sort the list using any column.}\label{t:pri}

\begin{center}
\begin{tabular}{r|p{.35\textwidth}|p{.35\textwidth}}
Column	& Definition	&Purpose \\\hline
TotDAT	& DAT sum over all diffs (over the last two years) modifying the file&	Shows the overall time in diffs that modify the file.\\
AvgDAT	& Geometric average: $\exp(average(\ln(DAT))$	& Geometric average over diffs: shows average diff time\\
totNormDAT & $\sum \sfrac{diffDAT}{diffNfiles}$ normalized DAT where the focus file takes only 1/files modified by a diff fraction of DAT	& For each diff attributes 1/files modified proportion of DAT to the file \\
avgNormDAT	& geometric average of the normalized DAT: $\exp(average(\ln(\sfrac{DAT}{nfiles})))$&	Geometric average over diffs for normalized DAT: shows average diff time attributable to file \\\hline
NDiff2Y	& Number of diffs modifying the file over past two years &Some files have been deprecated or less often changed in recent past, hence should be less a focus of refactoring \\\hline
n\SEVs	& Number of \SEV-related diffs modifying file	&Refactoring may improve quality\\
$\SEV_{level}$ & Highest severity over all SEVs \\\hline	
nAuthor	& Number of authors	& Files with many authors may indicate potential coordination problems \\
nDiffs	& Number of diffs (over the entire history) & \\	
nMnth	& number of months during which there were diffs (over the last two years)& Is the file being constantly modified? \\
diffsPerMonth	&&	Intensity of activity -- to compare recent and older files \\
fr &	Date of the first diff (within last two years)	& \\
to &	date of the last diff (within last two years) &	Has the file been changed recently? \\\hline
min and $\max_{cent}$ &	Lowest and highest file centrality over past two years	& Centrality trend\\
Avgdc &	average difference in file and developer centrality (over the last two years).& 	Developers with low centrality should have many more problems with the high-centrality files, so positive values for avgdc may indicate problems.\\
Pagerank and centrality &	represent code dependencies & These network measures represent to what extent central or isolated the file is in the dependency network \\\hline
knowLost &	percent of diffs by authors who are no longer with \Meta	& High fraction may suggest the need to strengthen ownership \\
AuthLeft& percent of authors who are no longer with \Meta & \\	\hline
sloc & lines of code &	to gauge the size of the file \\
complexity	& cyclomatic complexity &	to gauge complexity of the file (highly correlated with size) \\
nFilePerDiff &	average number of files modified by diffs touching the file	& Is the file relatively isolated or tied to other files? \\\hline
nTotAuth	& Total number of authors over the last two years & \\	
TopCochanged	& Other files that appear in at least 20\% of the diffs	& To show if the files is tightly logically connected to other files\\
\end{tabular}
\end{center}

\end{table*}

For many of the most central files in the spreadsheet, the feedback was ``yes, this is a really bad file and we are already improving (are planning to improve) it.'' 
We also got more nuanced responses, such as (slightly paraphrased) ``This one feels better than the other. It is touched a lot: any new logic in class X has to go through this file. But it is not really that complex: just a large function registering stages one by one.''
Or, ``This file has small complex parts that are not touched as much and very simple parts touched on day-to-day. Here is where analysis at the granularity of individual methods can help.`` 
Or, ``This is just an enum with ~400 entries and a map. Not sure how it ended up here. There definitely bigger enums (maybe not changed as frequently?)''
Regarding the bundling of files together we had a response: ``I would argue that the cpp file is more complex and central than the header... should be bundled as a single unit in all cases.'' However, in another case, a comment for \textrm{.h} file:
``Solid "bad" file, but not sure if we want to bundle with all of its cpps files.''

Regarding the persistence of improvements we had: ``Looks like this one has been simplified a lot over the last 1-2 years. Not sure if it still should be so high up in the list''. Also, ``Yes, this file is bad. People keep adding independent structs to it all the time. It is mainly bad for build speed. At some point I split a whole bunch of them into separate files, but it didn't cause a change in developers behaviors''

In summary, we found that the centrality metric could be improved if applied at a finer granularity of a function (thus, also solving the bundling question). Centrality also focuses on files that have system-wide impact, like registry of global constants or functions. Such registries might be, conceptually, very simple, thus incorporating code complexity into the centrality calculation might lead to a more generally applicable criteria. Some of the comments suggest that there may have been a regression since the previous improvement and 
that in some cases it may be important to change the development practices along with code improvement. The developer feedback did not explicitly address other metrics we provided, but some of the feedback supported the need to consider co-changed files that, like include files, developers sometimes consider as being a part of the main file and sometimes separately. We did not get the feedback that we missed important files, hence the filtering criteria using recent and numerous changes authors appear to be sound.

\section{Evaluating the impact of code improvement}\label{s:eval}

We worked with one of the suborganizations at \Meta that is constantly engaged in efforts to continuously improve its source code. Various goals for code improvement are set and tasks created every six months. We studied the impact of the work done in a single  six-month period. The tasks were grouped into four types of code improvements: dead code removal, Cyclomatic Complexity Number (CCN) decomposition, large class decomposition, and platformization as described in Table~\ref{t:retypes}. 

\begin{table*}[ht]

\caption{Description of the types of reengineering tasks.}\label{t:retypes}

\begin{center}
    \begin{tabular}{ p{.2\textwidth} | p{.3\textwidth} | p{.3\textwidth} }
    Type & Description & Risks \\\hline
    Dead code removal & Unused methods, data fields, classes, build targets, configs, feature flags, etc. &  ``dead'' actually used in emergency; 
``dead'' actually work in progress
 \\\hline
    Cyclomatic Complexity, CCN-based decomposition & The focus is on ways to reduce code complexity, and includes a variety of tasks such as refactoring fields or classes, unifying access to class fields, moving code to appropriate location, refactor a class into more logical units, organize functions and fields into more intuitive groups, and many other types of modifications & Due to the variety of the tasks combined under this label, the risk would vary with task type. \\\hline
Large class decomposition & The action is to do a subclass extraction, where tightly connected code in a ``super'' class is extracted and moved into a newly created class.  & To reduce risk typically done in three steps: (a) declaring a new class with all data dependencies
(b) moving method by method (or all methods at once) to a new class (retaining their body in the original file); (c) moving methods to a new class implementation file.
 \\\hline
    Platformization & The action is to replace Service Locator pattern using dependency inversion to enable a subsequent migration to a, for example, DAG  execution framework that can properly capture data dependencies and state mutation;
decompose god-objects and replace service-locator usage of god-objects with standard Dependency Injection
& Service Locator usage suffers from tight coupling and implicit data dependencies, making code and system state difficult to reason about\\
    \end{tabular}
\end{center}

\end{table*}

As outcome variables (see Table~\ref{t:measures}) we investigated  \SEV incidence, Diff Authoring Time (DAT), number of editing sessions, and how code complexity and centrality of the modified codebase has changed as a result of this reengineering. 
Our primary aim was to quantify the impact the reengineering had on quality (\SEVs), authoring time, and structural properties of the source code.
\vspace{-.1in}
\subsection{Method}
\vspace{-.05in}

As described in Section~\ref{s:proc}, the software changes are often a result of explicit tasks tracked in \Meta's task tracking system. The links between tasks and changes are carefully tracked and are routinely used to determine the effort needed to complete the tasks or, as in our case, to evaluate their impact. 
We started by identifying diffs associated with the code improvement tasks and then used these diffs to identify all modified files (we refer to them as ``reengineered files''). 

Diffs associated with the reengineering tasks are referred to as ``reengineering diffs''. 
We then identified all diffs modifying reengineered files during three months prior to and after the intervention (we refer to them as pre- and post-reengineering diffs correspondingly). 
We then obtain reengineered file properties before and after the reengineering and compare them as well as properties of pre- and post-reengineering diffs.

Since many of the files have been renamed, removed, or created by the reengineering diffs, we could not directly compare pre to post changes at the individual file level. Instead, we compare the distribution of code metrics over the entire set of reengineered filenames as they appeared both before and after the change. For example, metrics for deleted files were included to obtain pre-intervention distribution and metrics for created files in the post-intervention distribution.
For comparisons of process metrics, such as review time, we compared all diffs modifying at least one of the reengineered files over a three-month period immediately preceding the intervention and a three month period immediately following the intervention.
\begin{table*}[ht]
\caption{Description of the units and measures used.}\label{t:measures}

\begin{center}
    \begin{tabular}{c|c|p{.6\textwidth}}
    Unit & Metric & Explanation \\\hline
    diff & DAT & Diff Authoring Time starts with the first trace of activity that can be traced to a diff (e.g. first VS Code session) and ends with the last session. Work on sessions connected to other diffs is excluded. \\
    diff & Number of Sessions & The number of code editing sessions obtained from DAT \\
    diff & \SEV-trigger & whether or not the diff was determined to be a cause of a \SEV\\
    File & Call Graph Centrality & Katz centrality~\cite{katz1953new} of file-to-file call-graph (function definition to function invocation)\\
    File & Cyclomatic Complexity & The number of linearly independent paths ~\cite{mccabe76}\\    
    \end{tabular}
\end{center}

\end{table*}
We also obtained all diffs that modified at least one of the files involved in reengineering and marked which of these diffs caused a \SEV. One of key challenges was that a 
substantial number of the files have been renamed as is expected for the nature of the changes. 
We need to include past data on the renamed files to fully account for the codebase and activity 
before and after the code improvement takes place. To simplify that task, we always use the files 
current name, i.e., for past diffs we convert the filename at the time of the event to the current filename. We thus obtained full association between files and all diffs that modified these files in the past and in the future as well as all diffs that caused \SEVs. 

We use the code metrics database table to obtain the  cyclomatic complexity number (CCN) prior to the intervention and its latest value.  Finally, we obtain review time for each diff that modified at least one of the files involved in reengineering. We use Fisher’s exact test for contingency tables and regular (and paired where appropriate) t-test or paired Wilcoxon tests to compare the pre- and post-reengineering states.

Code centrality can be calculated on a combined network of code dependencies, author-to-changed-file, and co-changes (all files changed in a single commit). These networks may also be considered separately resulting in multiple centrality measures, such as call-graph centrality and co-change centrality.
We use Katz centrality~\cite{katz1953new} as it takes into account not just the number of neighbors but also their importance.
Comparing centrality before and after the intervention is complicated, because centrality depends on the entire graph, not just on the refactored files. The set of nodes will be different as some of the files will be added and others deleted (we use the same ID for renamed files). Furthermore, we use a normalized measure of centrality that forces all values to range from zero to 1, essentially dividing all values by the (unnormalized) centrality of the most central node in the entire graph. Even small changes in centrality of that node (even if it is not directly related to the refactored files), will affect centrality values for the remaining nodes.     

To compare the centrality of the reengineered files before and after
the intervention, we selected a random sample of not-reengineered  files that matched the distribution of the refactored set. This is a widely used case matching method that helps with estimating causal effects from observational data~\cite{stuart2010matching}. Specifically, it is desirable to replicate a randomized experiment as closely as possible by obtaining treated and control groups with similar covariate distributions. In other words, for each refactored file $f_r$, we find a file $f_{nr}$ of the same size and programming language that has similar centrality in the period before refactoring. We then compare 
$$c_{adj}(f_r,t_{pre})= c(f_r,t_{pre})/c(f_{nr},t_{pre})$$
with $c_{adj}(f_r,t_{post}),$
where $c(f,t)$ is the centrality of file $f$ at time $t$ and $t_{pre}, t_{post}$ are times before and after reengineering.

\subsection{Hypotheses}

The measures obtained as described in the previous section were based on certain hypotheses concerning expected impact of reengineering.

The primary objective of reengineering is to make it less likely that serious bugs are introduced. Hence:
\\\textbf{H1:} We expect the simplified code base will lead to a lower chance that a diff will cause a \SEV;

Additional advantages of the simplified codebase is that it is more clear where and how to make changes, resulting in a shorter authoring time:
\\\textbf{H2:} DAT will be lower after the code is rejuvenated. 
and
\\\textbf{H3} The simplified code base will streamline diffs requiring fewer code editing sessions.

Among the possible ways to reengineer the code (and that was considered in this particular reengineering effort) is to split the files based on distinct types of functionality. In effect, each resulting file will call functions in fewer other files (more on centrality). Hence: 
\\\textbf{H4:} We expect that the call-graph centrality will be lower after reengineering. 

One of the intended outcomes of reengineering is that the most complex parts of the code will be simplified. Hence:
\\\textbf{H5:} We expect that the cyclomatic complexity of the reengineered codebase will be lower;

\vspace{-.1in}
\subsection{Results for RQ3}

\textbf{RQ3: what is the impact of the code improvement
in terms of quality, productivity, lead time, and code centrality?}

The engineering teams took on over 1000 reengineering changes and modified over 1000 files. Table~\ref{t:counts} provides a summary of the number of changes per intervention/metric type. 

\begin{table*}[ht]
\caption{Numbers of reengineering diffs and affected files.}\label{t:counts}

\begin{center}
\begin{tabular}{llr}
Tasks& Number of diffs & Number of modified files\\\hline
CCN-driven decompositions & 897 & 789 \\
Large class decompositions&262&114\\
Dead code removal&193&316\\
Platformization&136&132\\

\end{tabular}
\end{center}

\end{table*}

To answer H1, the fraction of diffs modifying any of the reengineered files that caused an \SEV are in Table~\ref{t:sev}. 
The table only shows statistically significant decreases that occurred for files affected by dead code removal and for CCN-driven decompositions and shows very substantial improvements.

\begin{table*}[ht]
\caption{The odds ratio is 5.2 for dead code removal and 1.55 for decompositions showing 90\% and 55\% decrease in SEV-causing diffs modifying refactored files. The results are not statistically significant for platformization and large class decompositions. Instead of raw numbers, proportions of all (pre- and post-reengineering) diffs and triggers are shown}\label{t:sev}

\begin{center}
\begin{tabular}{llrr}

Type& Period & Proportion of Diffs (no Sev) during the period & Proportion of SEV-triggering Diffs \\\hline
Dead code removal&Before &57\%  &76\%\\
                 & After &43\%  &24\% \\
CCN-driven decomp.&Before&72\% & 80\%\\
                  &After &28\% & 20\%\\
\end{tabular}
\end{center}

\end{table*}

Substantial improvements in quality have been previously documented in other studies of reengineering, e.g.,~\cite{geppert2005refactoring}. Often this is a result of a more transparent codebase where it is easier not to overlook some unanticipated effects of a code change.

\begin{table}[ht]
\caption{The median DAT. The decrease is statistically significant (Mann-Whitney two sample  test)}\label{t:dat}

\begin{center}
\begin{tabular}{lr}
Type & Relative change \\\hline
Dead code removal & 0.59\\
CCN-driven  decomp. & 0.23 \\
Large class decomposition& 0.41\\
Platformization & 0.52\\
\end{tabular}
\end{center}

\end{table}

To answer H2, a simpler codebase should also make it easier and more straightforward to write the code. Indeed, Table~\ref{t:dat}  shows 41\% to 77\% reduction in DAT for all types of refactoring. We expect that this shorter authoring time will require fewer code editing sessions (and fewer time-consuming task context switches~\cite{meyer2014software}). 

To answer H3, the number of sessions shown in Table~\ref{t:nses} go down the most for Platformization and Large class decomposition. Dead code removal had the smallest but still sizeable effect.

\begin{table}[ht]
\caption{The median number of sessions went down from 19\% for dead code removal to 67\% for large class decomposition.}\label{t:nses}
\vspace{-.1in}
\begin{center}
\begin{tabular}{lr}
Type&  Relative change \\\hline
Dead code removal & 0.19 \\
CCN-driven  decompositions& 0.41\\
Large class decomposition & 0.67\\
Platformization & 0.65\\
\end{tabular}
\end{center}
\vspace{-.05in}
\end{table}

To answer H4, we find that adjusted centrality shows significant increases as shown in Table~\ref{t:centrality}. 
Several contributors to centrality, such as co-change and author-to-file relationships need to be collected over a 
significant time period, thus reengineering will not have an immediate effect. The second possibility is that reengineering included refactoring 
large classes that resulted in the increase in the number of files and, potentially in diffs that co-changed more files, thus increasing dependencies (and, in turn, centrality). 

In addition to process measures, such as \SEVs, DAT, and the number of code editing sessions, we expect the structure of the codebase to change as well. Specifically, (per H5) since many of the tasks were targeting cyclomatic complexity, we expect it to be lower after reengineering. As expected, we see the largest improvements (of 26\%) for CCN-driven decompositions, but even dead code removal and large class decompositions also lead to decreases as shown in Table~\ref{t:cc}.  
\begin{table}[ht]
\caption{The relative decrease in cyclomatic complexity. The decrease is statistically significant (Wilcoxon paired  test) except for platformization.}\label{t:cc}
\vspace{-.1in}
\begin{center}
\begin{tabular}{lr}
Type & Avg decrease in CycCmplex\\\hline
Dead code removal & 7\%	\\
CCN-driven decompositions & 26\% \\
Large class decomposition & 7\%\\
\end{tabular}
\end{center}
\vspace{-.05in}
\end{table}

\begin{table}[ht]
\caption{The relative increase in centrality. The increase  is statistically significant (Wilcoxon paired  test)}\label{t:centrality} 
\vspace{-.15in}
\begin{center}
\begin{tabular}{lr}
Type & Avg adj increase in centrality \\\hline
Dead code removal & 50\% \\
CCN-driven decompositions & 100\% \\
Large class decomposition & 113\% \\
Platformization & 95\% \\
\end{tabular}
\end{center}
\vspace{-.15in}
\end{table}

\vspace{-.05in}
\section{Discussion}\label{s:disc}
\vspace{-.05in}
It is not surprising that code quality receives significant attention and, at \Meta, multiple courses, tutorials, wiki pages are devoted to the topic. Similarly, software engineers like to create tools to support their work, including work on code improvements. Hence we see numerous specialized tools that are explicitly devoted to code quality that go beyond the traditional version control, issue tracking, code review, build, and deployment tooling. For example, the deadcode and data detection and removal tool has resulted in the deleted millions of lines of code and petabytes of data~\cite{shackleton2023dead}.

One, potentially unique aspect at \Meta is that many of the tools and practices are deployed across the entire organization and not, as is more typical, siloed within product units. While individual products have their specialized needs, leveraging common tools (and adapting them to serve these special needs) allows for a much larger number and variety of tools than a single product unit could support. It would be interesting to study the relative impact of various initiatives, such as considering code improvement in performance reviews, displaying progress in personal badges and gamification tools, on the intensity and quality of code improvement efforts. Despite the highly visible efforts to promote better engineering practices, most of the code improvement effort is done outside this umbrella. 

By discussing the question of targeting code improvement efforts, including major architectural changes to a large codebase, we arrived at over 20 distinct criteria that could be used to identify parts of code that, if improved, would yield largest dividends. The main idea is that actively changed code that is high risk and takes a lot of effort should be the focus of reengineering, rather than code with code smells that is peripheral, rarely changed, and is not involved in outages. While peripheral code may be easier to reengineer, the impact of such efforts is likely to be low.
While some are simple and obvious, like complexity and size, many were not previously considered in the literature for the task of prioritizing code rework. We also found that call-graph network properties, in many cases, provided partial indicators of problematic areas.

Our attempt to replicate prior code reengineering study found similar results: specifically, substantial improvements in quality and speed.
\vspace{-.1in}
\section{Threats to Validity}\label{s:limit}
\vspace{-.05in}
We first discuss issues related to generalizability as this is a case study at a single company. Next we consider construct validity and internal validity.
\vspace{-.15in}
\subsection{Generalizability}
\vspace{-.05in}
Generalizing conclusions from a case study in software engineering is complex because of a large number of potentially relevant context variables. The analyses in the present paper were performed at \Meta, and it is possible that outcomes would differ elsewhere. We cannot release our data, even in an anonymized format, because it would violate legally binding employee privacy agreements.  
However, our study involves a variety of products and developers. The software systems involved have millions of lines of code and 10's of thousands of developers who are both collocated and working at multiple locations across the world.
We also cover a wide range of domains from user facing social network products and virtual and augmented reality projects to software engineering infrastructure, such as calendar, task, and release engineering tooling. 
To determine if the findings generalize, a comparison with the results obtained elsewhere is needed. We find that many, but not all our findings have 
been consistent with prior work in industry settings, increasing the confidence that the results generalize. Some, for example a much larger fraction of perfective maintenance, 
may  be unique to \Meta or may be more common at present.
\vspace{-.1in}
\subsection{Construct Validity}
\vspace{-.05in}
In our study, we used the outcome measures that are commonly used at \Meta such as outages, diffs, code complexity, and diff duration metric (DAT). The DAT metric leverages existing tools in \Meta to get an accurate estimate of time spent working on a diff.  We also use more complex collaboration measures relying on software supply chains and centrality~\cite{bird2009putting,zimmermann2008predicting,zanetti2013categorizing,singh2010small,sureka2011using,wang2021characterizing}. 

We analyzed all files touched by reengineering diffs, but some of the modified files may not have been the actual targets of reengineering, but were changed together with the reengineered files.  We do not expect many such instances as the reengineering initiative was undertaken separately from the regular coding and maintenance activities. Organic reengineering effort, on the other hand, might be undertaken as part of regular coding activities and might include 
\vspace{-.1in}
\subsection{Internal Validity}
\vspace{-.05in}
We have used only most basic statistical tests and checked when needed if the assumptions on the distributions were reasonable. As is commonly necessary for statistical models in software 
engineering~\cite{Boehm-81,M08,Changes07,M10,M14}, we log-transformed variables with
highly skewed distributions.
\vspace{-.1in}
\section{Conclusion}\label{s:conc}
\vspace{-.05in}
Across \Meta and the broader software engineering community there is sparse data on how much code improvement activity happens, how it is organized or encouraged, how it is (or should be) prioritized, and what impact it has on key software engineering outcomes. 
We conduct a multifaceted analysis of the reengineering effort undertaken at \Meta, by doing a search for related practices, tools, and reward mechanisms. We also conduct a bottom-up approach to quantify the relative number of changes corresponding to different types of code improvement.  

Code improvement activities do take many forms, ranging from major quality initiatives that are well documented, to individual engineer driven actions that are less visible. In addition to courses, tutorials, extensive documentation, and regular scheduled code improvement activities, extensive tool support is provided to help track development activity and code quality metrics over time and also include various ways to engage developer 
via profile badges and even gamification of code improvement activities. We found that over 14\% of the changes were made explicitly for code improvement: substantially higher than previously reported 4\% in~\cite{mockus2000identifying}. We further discuss how to target strategically important parts of the codebase by producing a variety of indicators engineers could use to prioritize their work. 

Finally, we analyze the reengineered files to track the impact of the reengineering work over time. We observe several types of these targeted activities ranging from removal of dead code to sophisticated major restructuring aimed to achieve greater modularity by inverting dependencies. We found significant reductions in authoring time (DAT) and the number of coding sessions for the code targeted by the reengineering work, substantial reductions in \SEV incidence, and  reductions in code complexity. Surprisingly, call graph centrality did not decrease and further work is needed to determine if it is an expected outcome of code improvements or if the methodology of measuring changes in centrality should be improved.  

Our findings suggest that code improvement and reengineering practices need to be continuous in nature and supported by tools, as, for example, continuous build, to counteract potential issues that inevitably get introduced in active and rapid development.
We hope that further studies in other companies would shed light on which of our findings hold more broadly and which may be specific to \Meta.

\balance
\bibliographystyle{IEEEtran}
\bibliography{ref,audris}
\balance

\end{document}

%% file: PeterMacros.tex
\usepackage{xcolor}
\usepackage{xspace}

\PackageWarning{TODO:}{X}
\PackageWarning{TODO:}{X\%}

\usepackage{tcolorbox}